\documentclass[lettersize,journal]{IEEEtran}
\usepackage{amsmath,amsfonts}
\usepackage{algorithmic}
\usepackage{algorithm}
\usepackage{array}
\usepackage[caption=false,font=normalsize,labelfont=sf,textfont=sf]{subfig}
\usepackage{authblk}
\usepackage{textcomp}
\usepackage{stfloats}
\usepackage{url}
\usepackage{verbatim}
\usepackage{graphicx}
\usepackage{cite}
\usepackage{xcolor}
\usepackage{enumitem}
\usepackage{soul}

\begin{document}
\newpage
\thispagestyle{empty}

\noindent\begin{minipage}{\textwidth}
    {\Huge\textbf{IEEE Copyright Notice}} \\ 

    \vspace{2cm}
    \Large{\copyright 2023 IEEE. Personal use of this material is permitted. Permission from IEEE must be obtained for all other uses, in any current or future media, including reprinting/republishing this material for advertising or promotional purposes, creating new collective works, for resale or redistribution to servers or lists, or reuse of any copyrighted component of this work in other works.}
\end{minipage}

\vspace{2cm} 

\noindent\begin{minipage}{\textwidth}
    
    \LARGE{Accepted to be Published in: \textbf{IEEE Communications Magazine} \\ \\
    DOI: \textbf{10.1109/MCOM.2200084}}
    
\end{minipage}

\newpage
\title{\textcolor{red}{\small{This paper has been accepted to be published in IEEE Communications Magazine (Volume: 61, Issue: 3, March 2023). \\DOI: 10.1109/MCOM.2200084}} \\ [2ex] An Architecture for Provisioning In-Network Computing Enabled Slices for Holographic Applications in Next-Generation Networks}

\author[*]{Fatemeh Aghaaliakbari}
\author[**]{Zakaria Ait Hmitti}
\author[*]{Marsa Rayani}
\author[**]{Manel Gherari}
\author[*,+]{Roch H. Glitho}
\author[**]{Halima Elbiaze}
\author[**]{Wessam Ajib}

\affil[*]{CIISE, Concordia University, Montreal, QC, Canada}
\affil[+]{Computer Science Programme, University of Western Cape, Cape Town, South Africa}
\affil[**]{Université du Québec à Montréal (UQAM), Montreal, QC, Canada}




\maketitle

\begin{abstract}
Applications such as holographic concerts are now emerging. However, their provisioning remains highly challenging. Requirements such as high bandwidth and ultra-low latency are still very challenging for the current network infrastructure. In-network computing (INC) is an emerging paradigm that enables the distribution of computing tasks across the network instead of computing on servers outside the network. It aims at tackling these two challenges. This article advocates the use of the INC paradigm to tackle  holographic applications' high bandwidth and low latency challenges instead of the edge computing paradigm that has been used so far. Slicing brings flexibility to next-generation networks by enabling the deployment of applications/verticals with different requirements on the same network infrastructure.    
We propose an architecture that enables the provisioning of INC-enabled slices for holographic-type application deployment. The architecture is validated through a proof of concept and extensive simulations. Our experimental results show that INC significantly outperforms edge computing when it comes to these two key challenges. In addition, low jitter was maintained to preserve the hologram's stability.
\end{abstract}

\begin{IEEEkeywords}
Holographic applications, in-network computing (INC), slicing, applications/services provisioning, next-generation networks.
\end{IEEEkeywords}


\section{Introduction}

\IEEEPARstart{L}{et} us imagine a distributed concert at the Internet scale with audiences scattered around the globe, but with the perception that everybody is immersed in the concert with the performers on stage. To realize this vision, several cameras will surround the performers to capture multiple angles and views. The cameras' outputs will be merged, rendered into holograms, encoded, and streamed over the network. On the audience's end, the streams will be decoded and rendered by holographic displays. This concert belongs to an emerging group of applications known as holographic applications. Other examples are holographic learning and holographic surgery. They are still in their infancy and are usually operated today on a very limited scale due to their stringent requirements \cite{hologram,remote}.

The three key requirements of holographic applications that are not met by the current networking infrastructure are high bandwidth, ultra-low latency, and the ability to dynamically synchronize data streams \cite{hologram}.This paper focuses on the first two requirements, high bandwidth and ultra-low latency. They are extremely challenging to meet.
Novel mechanisms are therefore needed. For the high bandwidth, for instance, techniques are needed to reduce the network load. Edge computing \cite{Carla} enables computing close to end-users and has been used in the literature to tackle various aspects of provisioning holographic applications (e.g., \cite{remote, optimizing}). Unfortunately, the results are far from compelling when it comes to the two mentioned requirements.
As this paper shows, architectural components and experimental results are still scarce and unconvincing.
This paper takes a different standpoint and considers the use of in-network computing (INC) \cite{dumb}, a paradigm that enables flexible distribution of computation tasks across the network instead of computing on the servers or devices outside the network \cite{dumb}. 
It is fundamentally different from edge computing due to the fact that  
computation takes place inside the network instead of outside the  
network.
Our crucial motivation for selecting INC is that it aims at high bandwidth and ultra-low latency \cite{dumb}. Slicing \cite{slice} enables the co-existence of applications with different requirements in the same network. Holographic applications therefore need to be provisioned as slices since they will certainly co-exist with other applications in next-generation networks. Unlike the other slices, these slices will need to be INC-enabled. This paper proposes an architecture for provisioning INC-enabled slices for holographic applications in next-generation networks.
Our first contribution is a set of architectural components that enables INC-enabled slice provisioning for holographic applications. The second contribution is a set of concrete measurements showing that the proposed architecture outperforms the edge-based approaches for both high bandwidth and ultra-low latency. Furthermore, we demonstrate that our proposed approach is capable of keeping the jitter value within 15 ms, hence ensuring the hologram's stability \cite{itu}. We have used an experimental environment with P4, the programming language most widely used in INC settings. The holographic concert is used as a use case throughout the paper.

The next section provides background information on holographic applications, INC, and slicing. The third section critically reviews the state of the art, focusing on edge-based solutions. This is followed by the proposed architecture and its validation. Research directions are discussed last.

\section{On Holographic Applications, In-Network Computing, and Slicing}
\subsection{Holographic Applications}

Reference \cite{hologram} gives an overview. This subsection provides a synopsis. Holographic applications rely on transmission of and interactions with holographic data from remote locations across the network. In the holographic concert, for instance, remote performers will be projected as holograms to the remote audiences, who will be able to interact in real time with the performers. 

Holographic content is fundamentally different from 2D and 3D content. The viewer is active and can interact with holographic content but is passive when it comes to 2D/3D content. A 3D image comprises two 2D static views, while a hologram adds parallax. The parallax enables the viewer to interact with the image, and the image changes depending on the position from which it is being viewed. 
For transmission over the network, the critical function is transcoding. It consists of decoding, reformatting, and re-encoding to deliver better playback on the receiving end with heterogeneous display capabilities. It typically reduces the size of the hologram, and its implementation as INC will be considered in this paper.

The bandwidth requirement on the network can go up to 100 Gbps to 2 Tbps while it is just 1–5 Mbps for HD video transmission. Ultra-low latency is also required to avoid cybersickness, especially when a head-mounted display is used. In order to maintain the stability of the hologram, it is crucial to keep the jitter value within the range of 15 ms \cite{itu}. The synchronization of the massive number of streams originating from different sensors or objects at different angles also remains a crucial challenge in the network. However, this paper does not address it. 
\subsection{In-Network Computing}
INC comes with the vision of enabling computation inside the network as an alternative to restricting computation to servers outside the network \cite{dumb}, \cite{computer}, \cite{exhaustive}. Reference \cite{dumb} calls it a dumb idea whose time has come. The idea dates back to the active networks of the late 1990s but has finally started gaining momentum in the late 2010s/early 2020s due to the advances in data plane programmability. 
 
Software-defined networking (SDN), the first step towards data plane programmability, enables the separation between the control and data planes. However, SDN has several shortcomings \cite{exhaustive}. Although it allows the customization of the control plane, it offers only protocol-dependent actions in the data plane. Furthermore, it does not allow infield runtime re-programmability. The emerging programmable data plane devices tackle these issues and bring much more flexibility. They encompass switches and network interfaces and are directly programmable by users with languages such as P4\cite{exhaustive}. 


P4 is a protocol-independent packet processing language that enables data plane programming. 
Users can describe how the packets should be processed in a switching element by creating network forwarding capabilities through match/action tables. 
As a domain-specific programming language for network devices, P4 does not support complex operations. For instance, the transcoding function is a difficult procedure that P4 does not natively implement. To address this issue, P4 extensions may be added in two ways: by introducing new primitives or by using extern instances \cite{extern}. In our implementation, we used a free software transcoder written in C as an extern function that we instantiated in the switch core through a P4 program. It should be noted that the deployment of externs is switch-dependent, and we have deployed our P4 program on BMV2 \cite{bmv2}, the reference P4 switch. 
 
The key benefit expected from INC is performance gain (e.g., network load and latency reduction). Reference \cite{exhaustive} gives several examples. For example, two scenarios are considered in offloading media traffic to programmable switches. In the first scenario, the computation is done on relay servers outside the network, while in the second, it is done inside the network on a programmable switch. In the first case, the latency varies from 100 $\mu$s to 3 ms; in the second, it is negligible. 
 
Let us end this overview of INC by stressing that there are still many open issues. It is, for instance, not yet clear which type of application will benefit the most from INC \cite{dumb}, \cite{computer}. 
Furthermore, general mechanisms for deploying these functions are also needed. Moreover, it is important to keep in mind that the spare capacity available on programmable devices for INC is limited. 

\subsection{Slicing}
Network slicing enables the partitioning of a physical network into several virtual networks, with each virtual network customized and optimized for a specific type of application \cite{slice}. Slice embedding enables the mapping of these virtual networks onto the physical resources. It includes resource discovery and resource allocation. These virtual networks, or slices, are enabled by technologies such as SDN. 

The life cycle of a slice comprises four phases: design, orchestration, activation, run time assurance, and decommissioning \cite{slice}. A catalog of the building blocks of network slices is created during the design. Specific slices are selected during the second phase based on users’ requirements. Performance indicators (e.g., resource utilization) are continuously monitored and correlated with service quality (e.g., end-to-end latency). Network slices can be reconfigured if necessary to meet service-level agreements. 

\section{State Of The Art} 



To the best of our knowledge, no work has so far considered the use of INC to simultaneously tackle the high bandwidth and ultra-low latency challenges of provisioning holographic applications (or closely related applications). Andrus et al. \cite{ztp}, for instance, focus on the latency requirement. They suggest delegating the computational tasks of the video stream processing of security cameras to programmable switches to meet the low latency demand. In conventional video management systems, a video analytics server is responsible for detecting events and forwarding the video to the video storage. Instead, the authors propose utilizing the computational resource of the programmable switches. When an event (like object movement) is detected, the switches replicate the packet streams and then apply specified P4 processing and filtering to the replicated packet streams. It should be noted that no concrete measurement is presented to show to what extent the low latency requirement is met. 

A paradigm that has been widely used to tackle the high bandwidth and ultra-low latency challenges is certainly edge computing.
However, to the best of our knowledge, no edge-based approach simultaneously tackles both challenges. Reference \cite{edge} deals with high bandwidth, while references \cite{optimizing, SRFog} deal with ultra-low latency.

Bilal et al.\cite{edge} take aim at the high bandwidth challenge in interactive media and video streaming applications. They investigate the potentials and opportunities of using edge computing to tackle the high bandwidth issue. They study the average bandwidth consumed by a video streaming application. They also highlight the potential role of edge computing in transcoding from high resolution to low resolution. Unfortunately, no concrete architecture is proposed, and there is no validation.

Erfanian \cite{optimizing} addresses ultra-low latency for live video streaming to multiple audiences in different locations. The author proposes an architecture based on edge computing, SDN, and network functions virtualization (NFV) to deploy a set of virtual transcoder functions closer to the viewers. The architecture has three layers: the application/control layer, the network core layer, and the network edge layer. The architecture is very high-level, and there is no concrete implementation or performance evaluation.


Santos et al. \cite{SRFog} focus on latency.  
Accordingly, they propose an architecture based on fog computing (a variant of edge computing \cite{Carla}), source routing, and micro-service to deliver VR content in next-generation networks. The architecture addresses the current limitations of centralized clouds by bringing services closer to end users, resulting in reduced network latency. The performance evaluation shows that their proposal outperforms the cloud-based approaches.




It is worth mentioning that a few approaches other than edge computing have also been used to tackle the ultra-low latency problem.
In \cite{hologram}, for example, the authors analyze the effect of decentralization on latency and investigate four different SDN architectures for solving the problem.  The results demonstrate that the fully distributed architecture achieves the best results in terms of latency.

Edge-based approaches have indeed been used to tackle the two challenges we take aim at in this paper. However, none of the proposed solutions simultaneously tackles both requirements, as we do in this paper. The article that tackles the high bandwidth challenge (\cite{edge}) presents no concrete architecture or validation. One of the papers that tackles the latency challenge (\cite{optimizing}) proposes a sketchy architecture but no implementation. Regarding the second one (\cite{SRFog}), the focus is on comparing cloud-based approaches to edge-based ones. 

\section{The Proposed Architecture}
In this section, we first discuss the design goals of our architecture. An overview is provided next, followed by a detailed presentation of the slice embedding service, the key component of the architecture. The holographic concert use case is used as an illustration whenever appropriate.
\subsection{Design Goals}

As indicated in the title, our proposed architecture aims at providing components and procedures for the provisioning of INC-enabled slices for holographic applications. Its first design goal is that holographic application providers should be able to request slices with specific characteristics such as bandwidth, delay, and the number of participants in holographic concerts. The second is that the slice provider should be able to embed requested slices into the actual resources of the network. This requires, among other things, the monitoring of network resource usage.
We employ network slicing  to make possible that the architecture is able to serve simultaneously multiple applications providers (holographic application providers as well as other application providers) with each application having its own requirements and running on its own slice. The support of INC by the slices is yet another goal since it is the way bandwidth and latency requirements will be met. The last goal is that the architecture should be able to deal with heterogeneity.

\begin{figure}[!t]
\centering
\includegraphics[width=\linewidth]{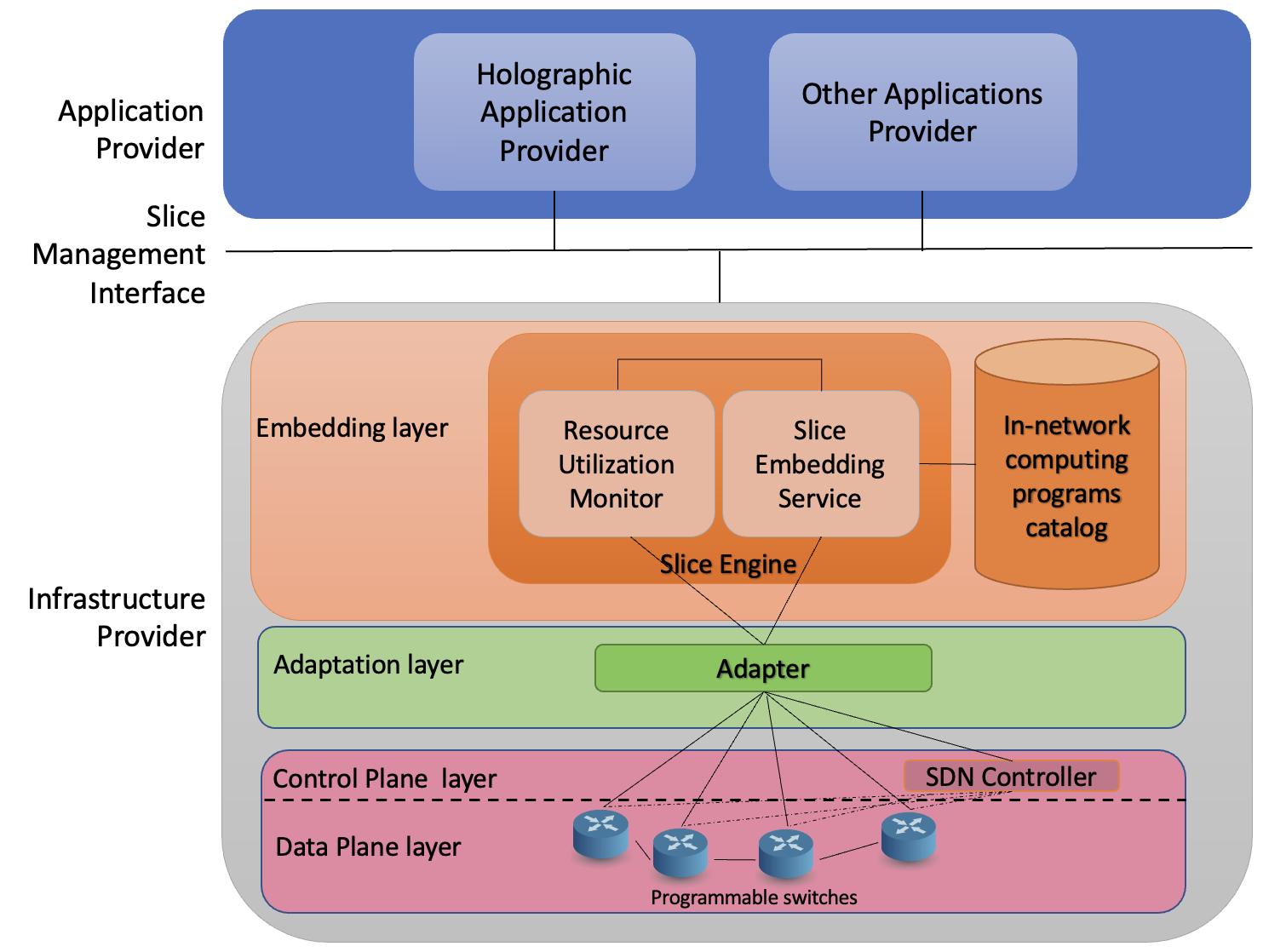}
\caption{Proposed Architecture.}
\label{fig:Arch}
\end{figure}

\subsection{Overview}
Figure~\ref{fig:Arch} depicts the proposed architecture. The functions enabling transmission and interactions with holographic data from remote locations across a network are the encoder, transcoder, decoder, and renderer.
The proposed architecture assumes that these functions have been developed using languages such as P4 and are compiled and ready for execution in programmable switches. The architecture rather focuses on the modules needed for provisioning INC-enabled slices that will make ultra-low latency and network load reduction possible.

Regarding Figure~\ref{fig:Arch}, the holographic application provider and the infrastructure provider interact via the slice management interface, which allows operations such as slice creation, updating, and deletion. Standards are now being defined for these interfaces, and any of these emerging standards could be used. 
In our use case, the application provider will be the holographic concert  
provider.
 
 The infrastructure has four layers, and we introduce these layers from bottom to top. The bottom layer is the data plane. It is made up of switches, and we assume that all the switches are programmable, meaning they are able to host and execute INC programs. The third layer is the control plane with the SDN controller. The second layer is the adapter layer. We envision next-generation networks as heterogeneous, and the goal of this layer is to mitigate this heterogeneity by offering a uniform interface to the slice embedding layer, which is the first layer. The slice embedding layer has two components: the INC program catalog and the slice engine. The INC programs are stored in the catalog, and the slice engine is made of two components: the resource utilization monitor and the slice embedding service. It should be noted that the INC program catalog comes in addition to the traditional slice building catalog previously discussed in this paper. The resource utilization collects the performance indicators in order to ensure that the requirements set by the application providers , such as holographic concert providers, are met. The slice embedding service carries out the actual embedding and INC-enabling. It is described below with the details of how it interacts with the other components of the architecture.

\subsection{Functioning of the Slice Embedding Service}

In this section, we use the holographic concert example to illustrate how the slice embedding service embeds a requested slice and makes it INC-enabled.


\begin{figure}[!t]
\centering
\includegraphics[width=\linewidth]{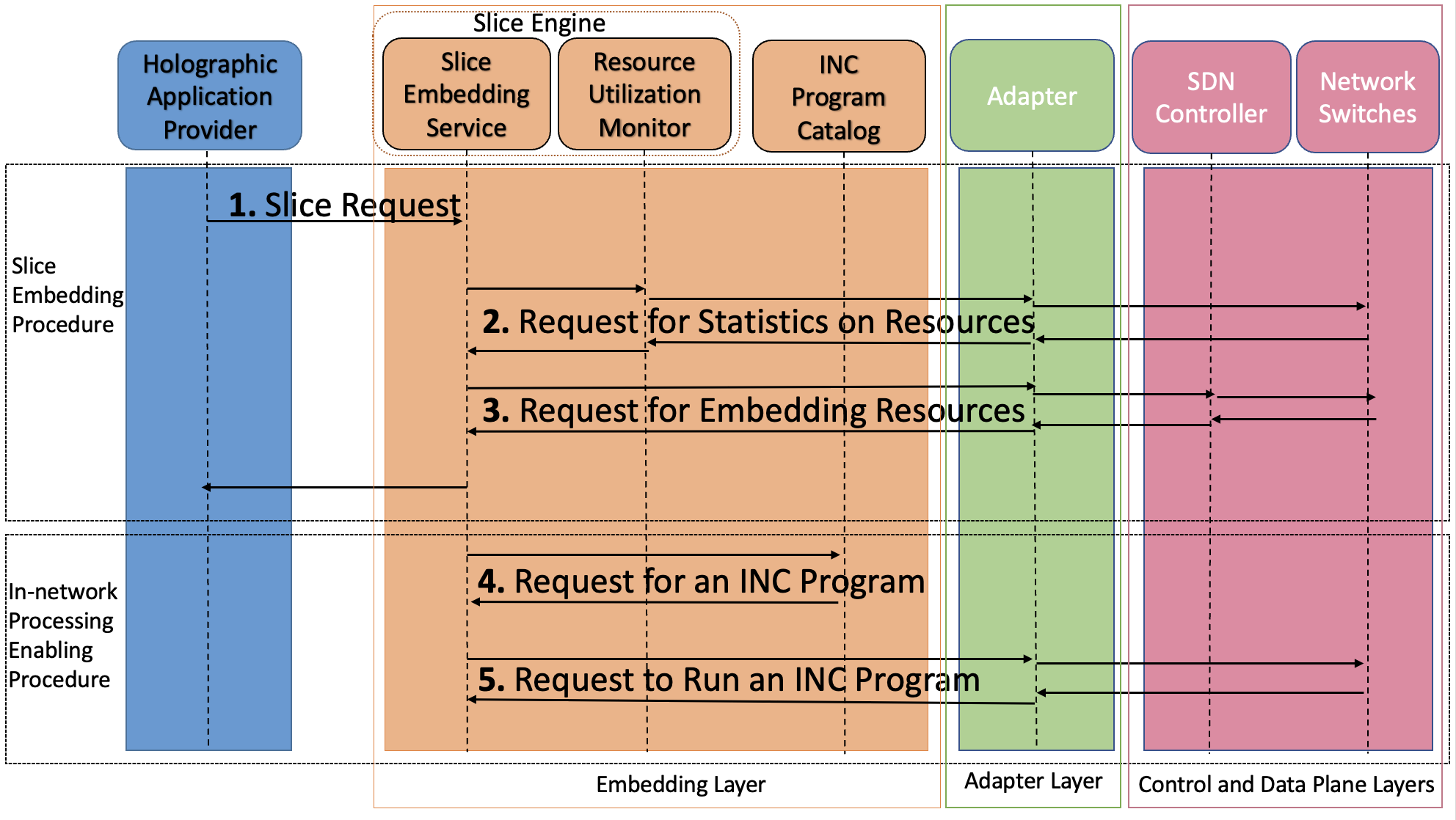}
\caption{Sequence Diagram of the Slice Embedding and INC Enabling Procedures.}
\label{fig:sequenece}
\end{figure}



Figure~\ref{fig:sequenece} demonstrates the interactions of the slice embedding service in detail. First, the holographic concert provider sends a request for a new slice creation to the slice embedding service,  which is shown in (1). 
This request consists of requirements, such as the required bandwidth and latency, the maximum number of remote attendees, and the localization of remote attendees in the case of a holographic concert.


Next, the slice embedding service requests to obtain performance indicators, such as CPU and links statistics, from the resource utilization monitor service (2). The utilization monitor service sends a request to the adapter to collect performance indicators from programmable devices directly or from SDN controllers to collect switch statistics. The adapter enables us to map our request to different SDN controllers and programmable switches to cover heterogeneous types of physical infrastructures. Then, the embedding algorithm decides to embed the slice based on the availability of infrastructure resources (3). 
There are several existing algorithms in the literature for embedding \cite{resource}, and any one of them could be used. 
Finally, the slice embedding service sends the notification of the slice creation to the holographic application provider. 

When the slice creation procedure is completed, the INC enabling procedure starts.  
The slice embedding service selects the INC program from the INC programs catalog, such as a transcoder (4). Then, the slice embedding service runs another algorithm to decide the placement of the INC program on the infrastructure switches (5). The algorithm decides to place the INC program on the infrastructure nodes based on objectives such as ultra-low latency and high bandwidth while considering the limited processing capability of the programmable devices. 
According to our investigation, there is no such placement algorithm in the literature. Also, other placement algorithms might not be applicable in the INC area. In our prototype, we have implemented a very simple algorithm described later.

\section{Proof of concept and performance evaluation}
\subsection{Experiment Environment} 

\begin{figure} [H]
\includegraphics [width=90mm, height=75mm]{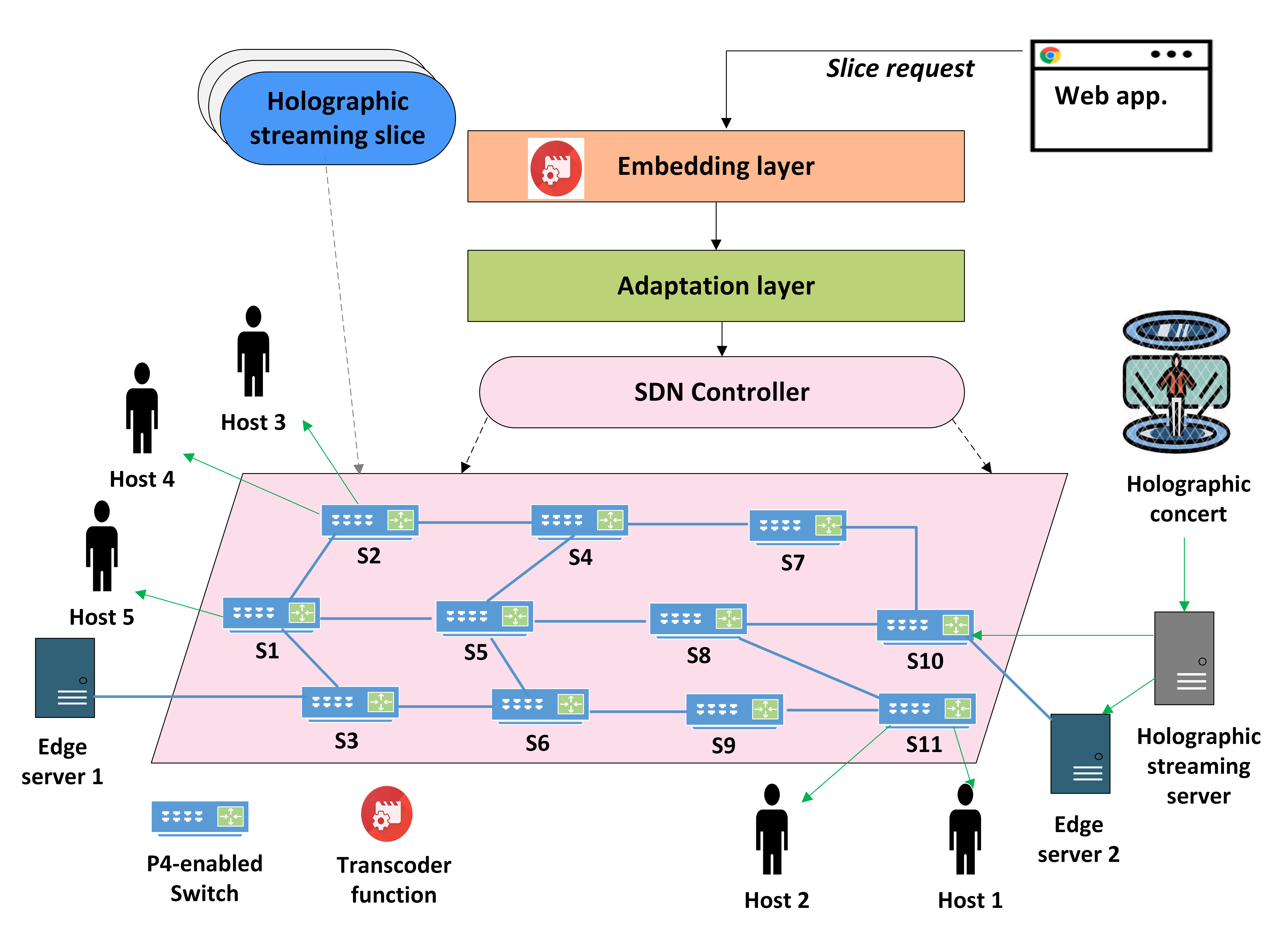}
   \caption{Network Setup} 
    \label{fig:SIMSetUP}
\end{figure}

We set up a P4-enabled SDN infrastructure using the Mininet emulator to i) validate the feasibility of our approach and ii) evaluate the performance of INC-enabled slicing for holographic applications. 
Figure \ref{fig:SIMSetUP} depicts the network configuration, which includes eleven P4 BMV2 switches [9] in the data plane managed by an SDN controller, an edge server, a holographic application provider, and five hosts representing the audiences.
We have developed and implemented the adaptation layer and the embedding layer modules to prototype our architecture described in Section IV. The adapter offers a RESTFul API to the embedding layer and maps the slice requests on the infrastructure. The \textit{SimpleResourceMonitor} module is developed to collect statistics on resource utilization, and the \textit{SimpleSliceManager} module is created to provide the slicing function.

\subsection{Simulation Scenario}  
We consider a simulation scenario where the holographic application provider generates a slice request to stream a holographic concert.  
Isolation among different slices is implemented using the \textit{Ethertype} field of the Ethernet header, which allows us to tag the packets for each slice.
Once the slice is created, the holographic application provider starts streaming the concert. The captured images are then merged, processed, and transmitted as a hologram through the network to the hosts. We assume that the holographic stream capturing the performers is already processed and merged, and only the transcoding function needs to be performed.
A hologram composed of 36000 frames is created after capturing live feeds of the performers and is streamed from the holographic application provider streaming server to the hosts. The frames represent images taken from different angles and distances to provide an immersive 6 DoF (degrees of freedom) for hosts.\\
\noindent
The hologram concert is streamed to the audience as follows:\\
\noindent 
\textit{- Transcoding at edge scenario 1}: The streams are transmitted to edge server 1 through the path (S10–S8–S5–S6–S3). The transcoded streams are then sent to hosts 1–2 via the path (S3–S6–S9–S11), to hosts 3–4 via the path (S3–S1–S2), and to host 5 via the path (S3–S1).\\
\noindent
\textit{- Transcoding at edge scenario 2}: The streams are transmitted to edge server 2, which is connected to the streaming server. The transcoded streams are then sent to hosts 1–2 via the path (S10–S8–S11), to hosts 3–4 via the path (S10–S7–S4–S2), and to host 5 via the path (S10–S8–S5–S1).\\
\noindent
\textit{- Transcoding at hosts}: The streams arrive at the hosts via the same route as in \textit{ Transcoding at edge scenario 2}. Also, the main difference to edge scenario 2 is that the packets do not move across edge server 2.

\noindent
\textit{- Transcoding in network devices near the audience}: We consider that three transcoders are placed in switches S11, S1, and S2. The streams follow the path (S10–S8–S11) to hosts 1–2, the path (S10–S8–S5–S1) to host 5, and the path (S10–S7–S4–S2) to hosts 3–4.\\
\noindent
\textit{- Transcoding in network devices near streaming server}: We place the transcoder in S10. The streams follow the path (S10–S8–S11) to hosts 1–2, the path (S10–S8–S5–S1) to host 5, and the path (S10–S7–S4–S2) to hosts 3–4.\\
The reader should note that edge scenario 1 and edge scenario 2 are the most widely deployed edge scenarios in practice. The simulations are run on a machine with a dual 2X8- Core 2.50GHz Intel Xeon CPU E5-2450v2 and 40GB of memory. The network switches have equal CPU capacity and 12 Mbps interconnecting links. In order to evaluate the effects of INC, we have different scenarios addressing two points of view: 1) the slicing is INC-enabled or not,  and 2) transcoding is delegated to either the edge servers, the hosts, or the network devices.
Network load is measured because we aim to tackle the high bandwidth challenge by reducing the network load. \\
The average end-to-end latency is calculated as the mean latency of all the packets transmitted during the holographic stream. The average network load is calculated as the mean used bandwidth per link divided by the mean bandwidth per link. The jitter is calculated as the mean latency variation between every two consecutive packets.
\subsection{Results}
\textit{Slice Creation Time Analysis}\\
We follow two scenarios to investigate the effect of enabling INC in slice creation. First, we create one INC-enabled slice using the transcoder on the programmable switches. Second, we create another slice without INC and have a transcoding program on the hosts. 
Slice creation time is defined by the elapses from the slice request until the time the slice is available for use. This process entails creating all of the slice's elements sequentially and grouping those resource chunks into the resulting slice. Slice creation time is obtained by running the scenarios ten times with a $95\%$ confidence level. Thus, INC-enabled slice creation takes $409.32 \pm 9.88 ms$, while the non-INC-enabled process takes $344.83 \pm 8.84 ms$. This can be explained by the additional steps (4 and 5) in Figure \ref{fig:sequenece} needed for INC-enabled slice creation. 
The P4 program that enables INC calls the transcoder extern function and implements an extra match/action table to manage the transcoding. The reader should note that the slice creation time has no impact on the holographic streaming phase because it takes place before the holographic concert starts streaming to the audience.

\textit{Latency and Jitter Analysis}\\
This section analyzes the latency and  jitter while placing the transcoder at the edge server, at the host, and at programmable network devices.

\begin{figure}[h]
 \centering
  \includegraphics[width=1\linewidth]{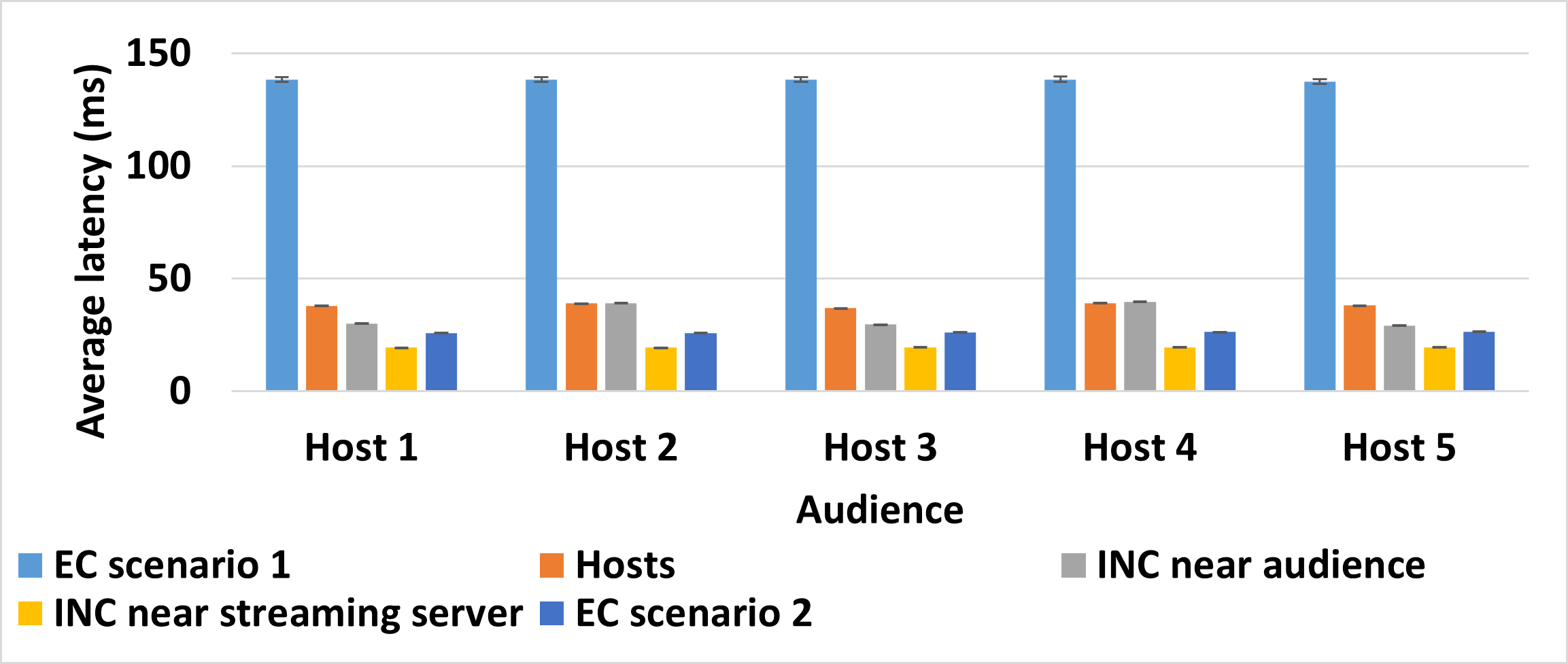}
   \caption{Average latency for each holographic stream}
    \label{fig:latency}
\end{figure}

As Figure~\ref{fig:latency} illustrates, transcoding on the edge server (\textit{EC scenario1}) results in high latency for all hosts compared to the other scenarios. This is a consequence of transmitting the original hologram data all the way to the edge to be transcoded and then sending it to the audience. Furthermore, transcoding on hosts significantly reduces the latency compared to the edge because we gain time wasted on waiting for edge responses.

Regarding the Figure~\ref{fig:latency}, for transcoding in the network near the audience, we notice a reduced latency for hosts 1, 3, and 5 compared to transcoding on hosts; however, hosts 2 and 4 show similar results for transcoding near the audience and on the hosts. This is because, in our implementation, holographic concert streams are sent in parallel through the shortest path to different hosts. Specifically, when we have two hosts connected to one switch, the streaming server sends each packet to the hosts sequentially. For example, for hosts 1 and 2 connected to switch 1, host 1 receives each packet first, followed by host 2. Hence, host 2 experiences more latency than host 1. 
Although placing the transcoder near \textit{the holographic streaming server} generates the best results, INC near the streaming server performs slightly better than \textit{EC scenario2}. This is due to the extra connection between the edge and the streaming server.

\begin{figure} [H]
  \centering
    \includegraphics[width=1\linewidth]{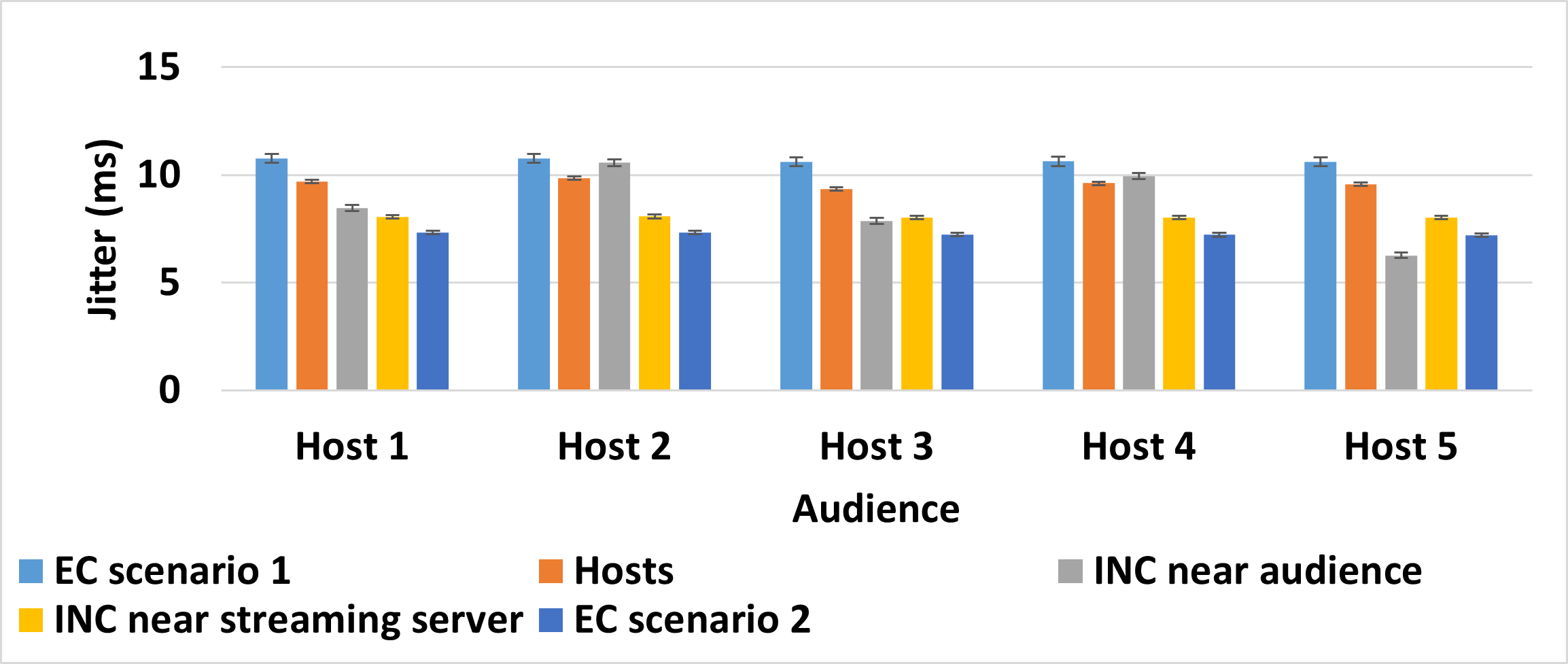}
    \caption{Jitter for each holographic stream}
    \label{fig:jitter}
\end{figure}

Another important requirement for streaming holograms is the jitter. Ideally, the jitter values should not exceed 15 ms \cite{itu} to avoid the hologram shaking during a live stream. Figure ~\ref{fig:jitter} depicts the jitter for different scenarios, which clearly demonstrates that the aforementioned constraint is satisfied in all scenarios. The difference in the jitter's values from one host to another is explained by the fact that the number of hops between the \textit{holographic streaming server} and each host differs, as illustrated in Figure~\ref{fig:SIMSetUP}. 

Overall, one important factor in reducing the latency is the placement of both the edge and the INC near the holographic streaming server. Nonetheless, INC slightly outperforms edge in reducing latency by transcoding along the path to hosts.\\
 
\begin{figure} [H]
    \centering
    \includegraphics[width=1\linewidth]{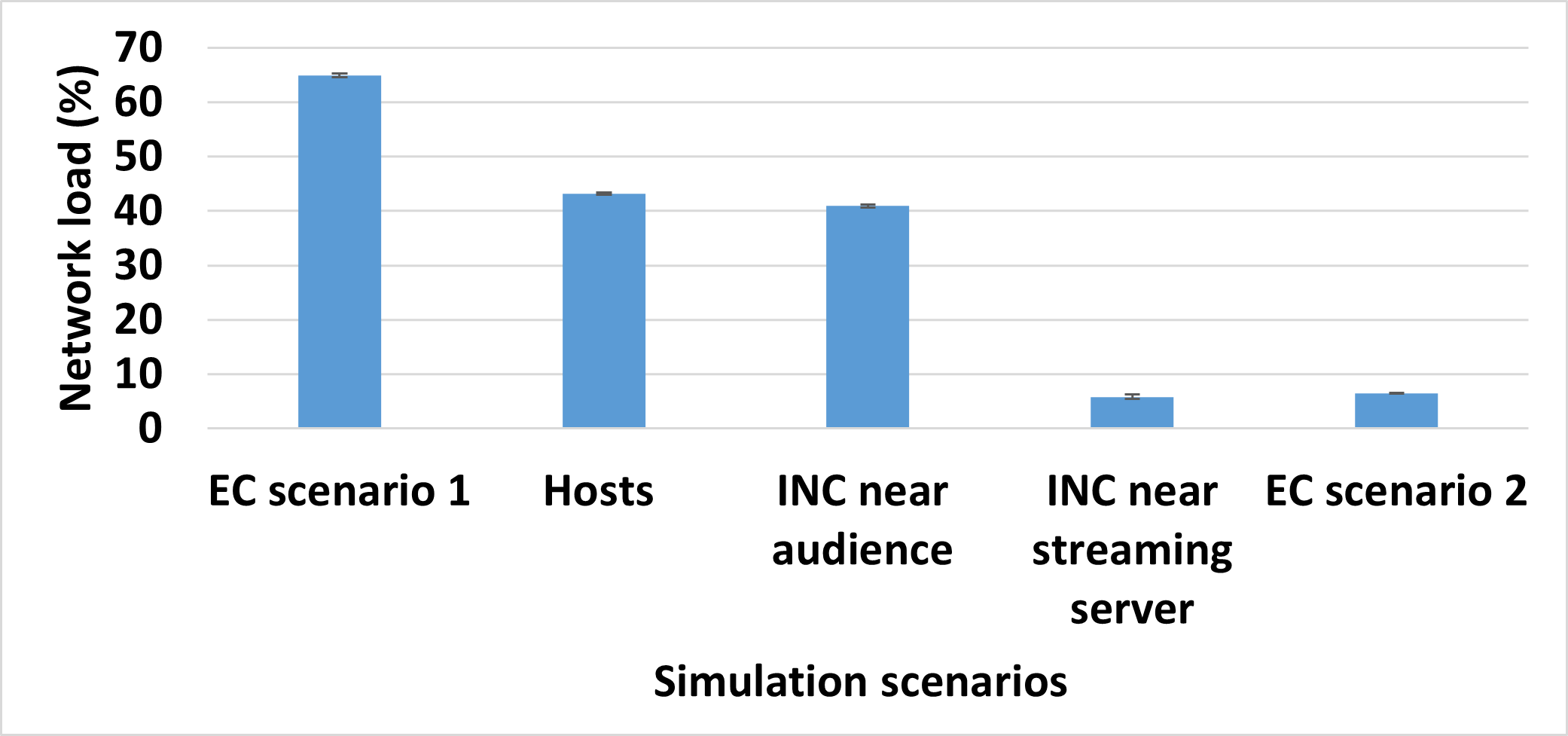}
    \caption{Network load for each scenario}
    \label{fig:networkload}
\end{figure}

\textit{Network Load Analysis}\\
Transmitting holographic data in their original heavy format can result in significant network congestion, as shown by Figure~\ref{fig:networkload}. Thus, in reducing the network load, INC outperforms both EC and hosts. Indeed, transcoding the hologram stream along the path and transmitting it in a smaller size reduces network load while maintaining high quality. It is essential to highlight that transcoding the hologram near where its data are generated offers the best results, maximizing bandwidth gain and reducing traffic load by more than 50\% compared to EC scenario 1.

\section{Research Directions} 

Several interesting research directions are worth pursuing. An example is the INC program placement problem. In this paper, the transcoder program was experimentally placed on different switches. Algorithms need to be developed in order to find the optimal placement. Objectives might be network load reduction, ultra-low latency, or both, and the constraints will include the capacity of the switches. The problem is actually more general. In addition to transcoders, other functions such as renderers and compressors could be considered. And in addition to the switches, hardware accelerators and network interface cards could also be considered for hosting the functions. Yet another example of a research direction is the design and validation of more general architectures that integrate not only INC, as done in this paper, but also other paradigms such as edge computing. Edge will certainly offer more computing capacity and could be used to host the functions that are not deployable as INC functions due to capacity limitations in the network. A third and last example of a direction is the management and orchestration of the slices. Research on the management and orchestration of slices has so far focused on slices that are not INC-enabled. However, INC-enabled slices bring a wealth of new challenges that need to be tackled.


\section{Biography Section}
\vspace*{-30mm}
\begin{IEEEbiography}[{\includegraphics[width=1in,height=1.25in,clip,keepaspectratio]{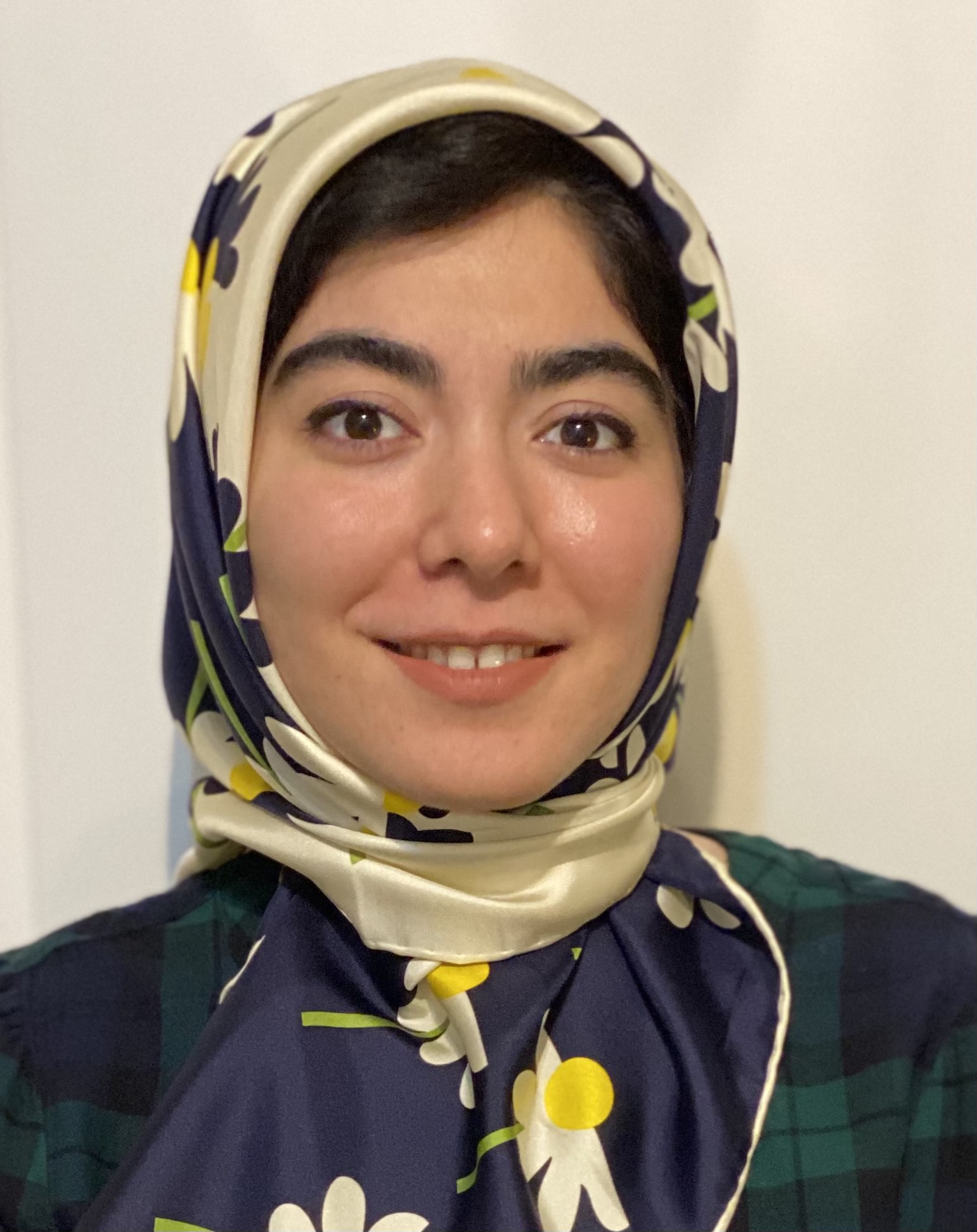}}]{Fatemeh Aghaaliakbari}
earned her Master’s Degree in Computer Engineering from Sharif University of Technology, Tehran, Iran in 2014. She worked as a Network Engineer at the Mobile Communication Company of Iran (MCI) for six years. She is currently a Ph.D. student in Information System Engineering at Concordia University, Montréal, Canada. Her research interests are Cloud Computing, Networked Systems, Virtualization, and Resource Management.
\end{IEEEbiography}
\vskip -4pt plus -1 fil
\vspace*{-9mm}
\begin{IEEEbiography}[{\includegraphics[width=1in,height=1.25in,clip,keepaspectratio]{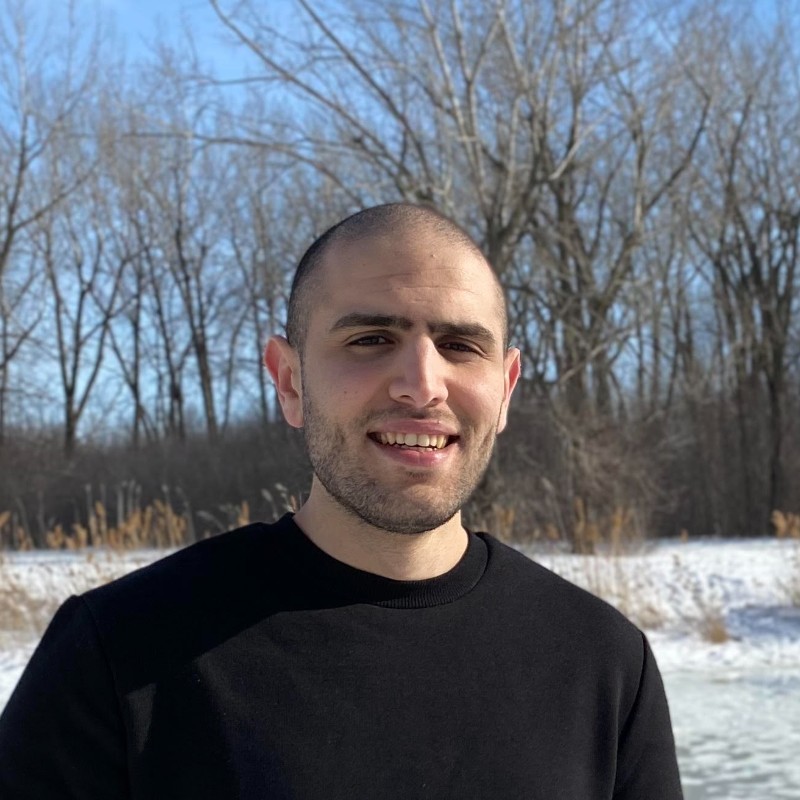}}]{Zakaria Ait Hmitti}
received the B.Sc. degree in mathematical and computer science from the Cadi Ayyad University, Marrakech, Morocco, in 2018, and the M.Sc. degree in internet of things and mobile services from the École Nationale Supérieure d’Informatique et d’Analyse des Systèmes, Rabat, Morocco, in 2020. He is currently a M.Sc student in the Université du Québec à Montréal. His research interests are P4, programmable data planes, software defined networking and in-network computing.
\end{IEEEbiography}
\vskip -4pt plus -1 fil
\vspace*{-9mm}
\begin{IEEEbiography}[{\includegraphics[width=1in,height=1.25in,clip,keepaspectratio]{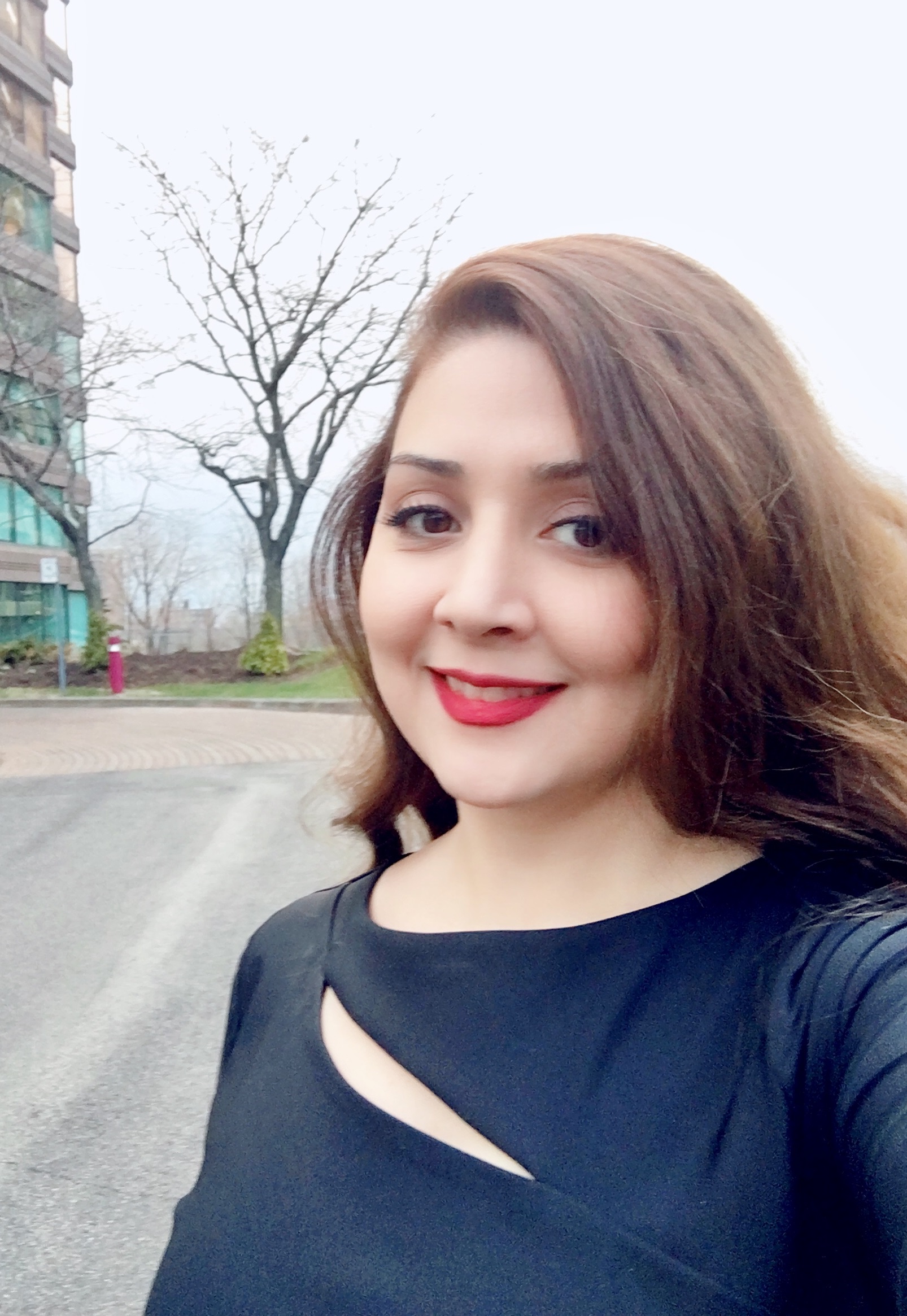}}]{Marsa Rayani} (Student Member, IEEE) is a Ph.D. student in Information System Engineering at Concordia University, Montréal, Canada since 2016, working on 5G, resource allocation, algorithm and design, Information Centric Networking (ICN) and Content Delivery Networking (CDN). She received her Bachelors in Computer Software Engineering in Iran in 2008 and her Masters in Computer Science, Distributed Computing from University Putra Malaysia.
\end{IEEEbiography}
\vskip -4pt plus -1 fil
\vspace*{-9mm}
\begin{IEEEbiography}[{\includegraphics[width=1in,height=1.25in,clip,keepaspectratio]{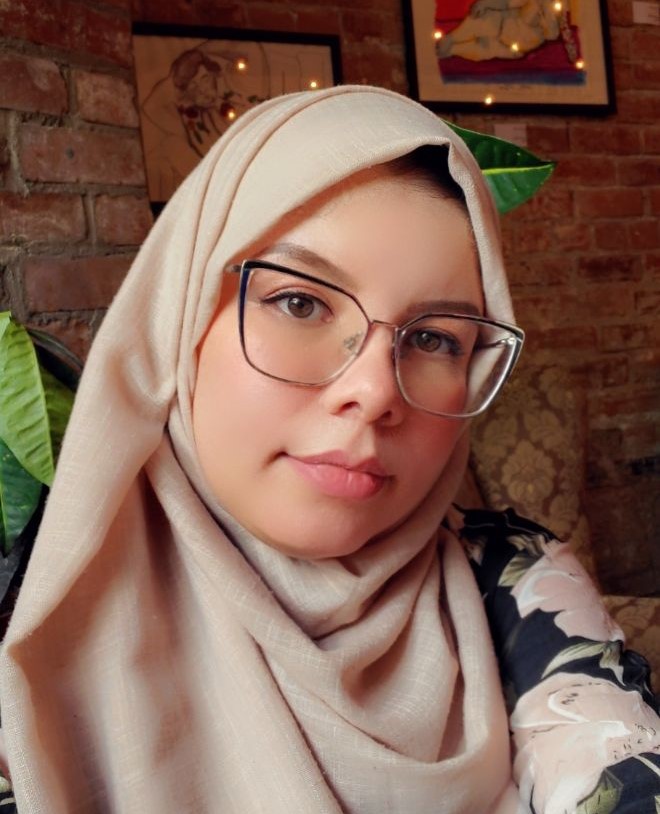}}]{Manel Gherari} earned her Ph.D. degree in Computer Science “Information System” from the University of Tebessa, Algeria in 2016. She received her Master's Degree in Software Engineering from the University of Souk-Ahras, Algeria in 2013. She worked as an assistant professor at the University of Annaba, Algeria for three years. Currently she is a Ph.D. student in network computing in the Université du Québec à Montréal (UQAM), Montréal. Her research interests are In-Network Computing, Internet of Things, Cloud Computing, and Network Virtualization.
\end{IEEEbiography}
\vskip -4pt plus -1 fil
\vspace*{-9mm}
\begin{IEEEbiography}[{\includegraphics[width=1in,height=1.25in,clip,keepaspectratio]{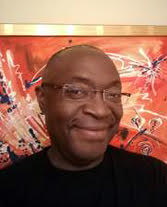}}]{Roch H. Glitho} is a Full Professor at Concordia University where he holds the Ericsson/ENCQOR Industrial Research Chair in Cloud/Edge for 5G and Beyond. He has held a Canada Research Chair from 2010-2020, and prior to joining academia in 2010 he has held several senior technical positions at Ericsson. He is also a Professor Extraordinaire at the University of Western Cape, South Africa.
\end{IEEEbiography}
\vskip -4pt plus -1 fil
\vspace*{-9mm}
\begin{IEEEbiography}[{\includegraphics[width=1in,height=1.25in,clip,keepaspectratio]{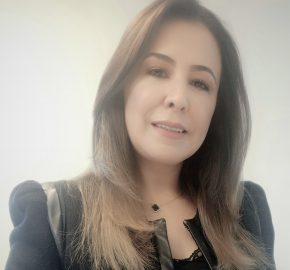}}]{Halima Elbiaze}
(M'03) received a Ph.D. degree in Computer Science from the Télécom Sud-Paris, France, in 2002. Since 2003, she has been with the Department of Computer Science, Université du Québec à Montréal, Montr\'eal, QC, Canada, where she is currently an Full Professor. Her current research interests include network performance evaluation, traffic engineering, and quality of service management in next generation networks. 
\end{IEEEbiography}
\vskip -4pt plus -1 fil
\vspace*{-9mm}
\begin{IEEEbiography}[{\includegraphics[width=1in,height=1.25in,clip,keepaspectratio]{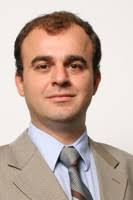}}]{Wessam Ajib} (Senior Member, IEEE) received his Engineer Diploma from Institut National Polytechnique de Grenoble, France, in 1996, master’s and Ph.D. degrees from the École Nationale Supérieure des Télécommunication, Paris, France, in 1997 and 2000, respectively. After working four years with Nortel Networks, Ottawa, ON, Canada, and one year as a Postdoctoral Fellow at the École Polytechnique de Montréal, QC, Canada, he joined the Universite du Quebec, Montreal, Canada, in 2005 where he is currently a full professor in computer sciences. His research interests include wireless communication and networks, machine learning algorithms for wireless networks, and resource allocation in 5G and 6G. 
\end{IEEEbiography}

\vfill

\end{document}